# Exploring the current applications and potential of extended reality for environmental sustainability in manufacturing

Huizhong Cao[1]*, Henrik Söderlund[1]*, Mélanie Derspeisse[1], Björn Johansson[1]

[1] Chalmers University of Technology, 41296, Gothenburg, Sweden

**Abstract**

In response to the transformation towards Industry 5.0, there is a growing call for manufacturing systems that prioritize environmental sustainability, alongside the emerging application of digital tools. Extended Reality (XR) — including Virtual Reality (VR), Augmented Reality (AR) and Mixed Reality (MR) — is one of the technologies identified as an enabler for Industry 5.0. XR could potentially also be a driver for more sustainable manufacturing: however, its potential environmental benefits have received limited attention. This paper aims to explore the current manufacturing applications and research within the field of XR technology connected to the environmental sustainability principle. The objectives of this paper are two-fold: (1) Identify the currently explored use cases of XR technology in literature and research, addressing environmental sustainability in manufacturing; (2) Provide guidance and references for industry and companies to use cases, toolboxes, methodologies, and workflows for implementing XR in environmental sustainable manufacturing practices. Based on the categorization of sustainability indicators, developed by the National Institute of Standards and Technology (NIST), the authors analyzed and mapped the current literature, with criteria of pragmatic XR use cases for manufacturing. The exploration resulted in a mapping of the current applications and use cases of XR technology within manufacturing that has the potential to drive environmental sustainability. The results are presented as stated use-cases with reference to the literature, contributing as guidance and inspiration for future researchers or implementations in industry, using XR as a driver for environmental sustainability. Furthermore, the authors open up the discussion for future work and research to increase the attention of XR as a driver for environmental sustainability.

## 1 INTRODUCTION

In order to stay competitive in a market with a growing demand for sustainable products and services, manufacturing companies are embracing smart and sustainable manufacturing practices as the core drivers for their Industry 5.0 transformational shift [1]. Extended reality (XR) as an enabler for this industry shift has previously been explored and discussed by multiple researchers, defining its applications in the manufacturing areas of the future. Previous research has proven a strong relationship between XR technologies in manufacturing together with economical and social sustainability, as XR provides great decision support and enables human centric solutions. However, the effects that XR technologies potentially have on the environmental aspects of sustainability are more seldomly discussed and to some extent still uncertain [2] as the potential benefit to environmental sustainability that XR brings is usually only indirectly proven.

### 1.1 Extended Reality

Extended Reality (XR) is the umbrella term for technologies that act as interfaces between the real and virtual worlds. The term includes Virtual Reality (VR), Mixed Reality (MR) and Augmented Reality (AR) which all offer different levels of immersion and level of interactions with digital objects between the real and virtual world [3].

VR applications completely immerse the user in a virtual world, usually using a head-mounted display (HMD) or projections that encapsulate the users with a full visual experience of a virtual world. By tracking the motions and position of the user, the motions and position can be mimicked in the virtual world giving the perception of full immersion [3]. For applications where full virtual immersion is not needed, MR technology can help us to mix and overlay 3D elements from both the real and virtual world in the same experience and environment. Usually, this is done via an HMD in combination with visual inputs from the real-world using cameras or AR headsets with a transparent display [3]. Finally, AR technology allows us to remain in the real world while overlaying digital information and augmented visuals on top of our perception of the real world. Examples of such technologies could include AR headsets, smart glasses, smartphone applications or projections [3].

The use of XR technology is often associated with industries such as gaming or entertainment. However, XR technology has also been widely implemented in manufacturing and has been identified as one of the enabling technologies for Industry 5.0 [4].

## 1.2 Environmental sustainability in manufacturing

Sustainable manufacturing has been defined in literature as a manufacturers ability to continue to assure demand and improve quality of human life while satisfying economical, societal and environmental objectives, not compromising future generations to meet their needs [5]. However, as a manufacturer it might not clear how to actually measure, and therefore achieve, sustainable manufacturing [6]. One possible way to measure the sustainability impact of a manufacturing system is by using the sustainability indicators developed by the US National Institute of Standards (NIST). NIST has identified a total of 213 indicators, that can be used to measure the sustainability in a manufacturing process, categorized under these three pillars of sustainability, environmental-, social- and economic sustainability [6][7]. Out of these, 77 indicators have been identified under the pillar of environmental sustainability and primarily focused on minimizing emissions, resource consumption, and waste as a result of the manufacturing processes [6]. In Fig. 1 an illustration of the areas covered by the 77 environmental indicators proposed by NIST can be seen.

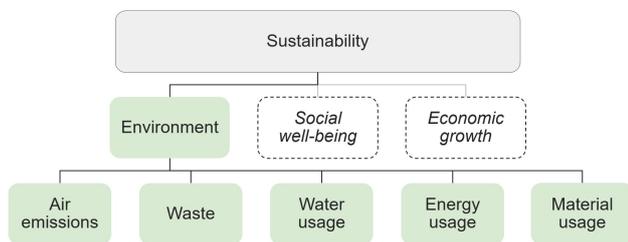

Fig. 1: Illustration of areas covered by the environmental indicators defined by NIST, adapted from [8]

By considering the sustainability indicators, identified and categorized by NIST, manufacturers, and researchers are able to measure certain actions and activities in their manufacturing system to understand how they affect environmental sustainability and furthermore how they can improve their sustainable manufacturing practices [7].

## 1.3 Research questions

In recent years we have seen an increase of literature exploring XR technologies in manufacturing as well as actual real case implementations in industry. At the same time some efforts have been put towards documenting the relationship between various XR tools and sustainable manufacturing [21]. However, there is still a gap in the knowledge that merges the broader concepts of XR, its systemization and use cases in manufacturing and environmental sustainability.

This paper aims to provide a better understanding of the direct or indirect contributions that XR technologies potentially can have on environmental sustainability in various manufacturing use cases. By mapping existing manufacturing use cases towards proven environmental impacts from existing literature, the authors hope to provide directions of future XR development for industry and researchers. By providing clear examples and closing this gap, acceleration towards sustainable manufacturing and Industry 5.0 using XR technology is aspired. The research questions for the paper are:

**RQ1**: How has sustainability been addressed in research efforts on XR for manufacturing applications?

**RQ2**: What environmental benefits have been obtained in use cases of XR in manufacturing?

## 2 METHOD

The method developed for this paper categorizes the XR use cases according to the sustainability indicators. By identifying relations between the two, such direct/indirect and with/without evidence, to sort the literature in order to find the correlations and research gap in strengthening the relationship between XR and environmental sustainability.

### 2.1 Literature search, scope and inclusion of papers

As shown in Fig. 2 to meet the objective requirements, we devised the search strings in the database Scopus, with the keywords "sustaina*", "XR" and its extension, "Production" OR "Manufac*" to look for the nonbiased trend in the XR development for sustainability. The first selection was based on the criteria of the year of publication, subject area, and language. After the first selection, the use cases derived from the selected papers were further screened by their conclusions and 30 papers with clear use cases in sustainable manufacturing are processed and fully studied within the literature review.

### 2.2 Analysis

In order for the authors to conduct the analysis and to answer the posed research questions from section 1, a spreadsheet was created to capture data of the examined papers used for the analysis. The spreadsheet recorded generic meta data of each paper such as year of publication, country of origin etc. but also specific data points used in the analysis of use cases. Detailed information about the paper, including the hardware, software used, sustainability dimensions, evidence, limitations, and reflection, was collected and examined. The criteria recorded as part of the analysis were the following: (i) The application and use of XR technology proposed by the author, (ii) The corresponding sustainability pillar(s) impacted by the use case, (iii) Direct/Indirect impact on environmental sustainability.

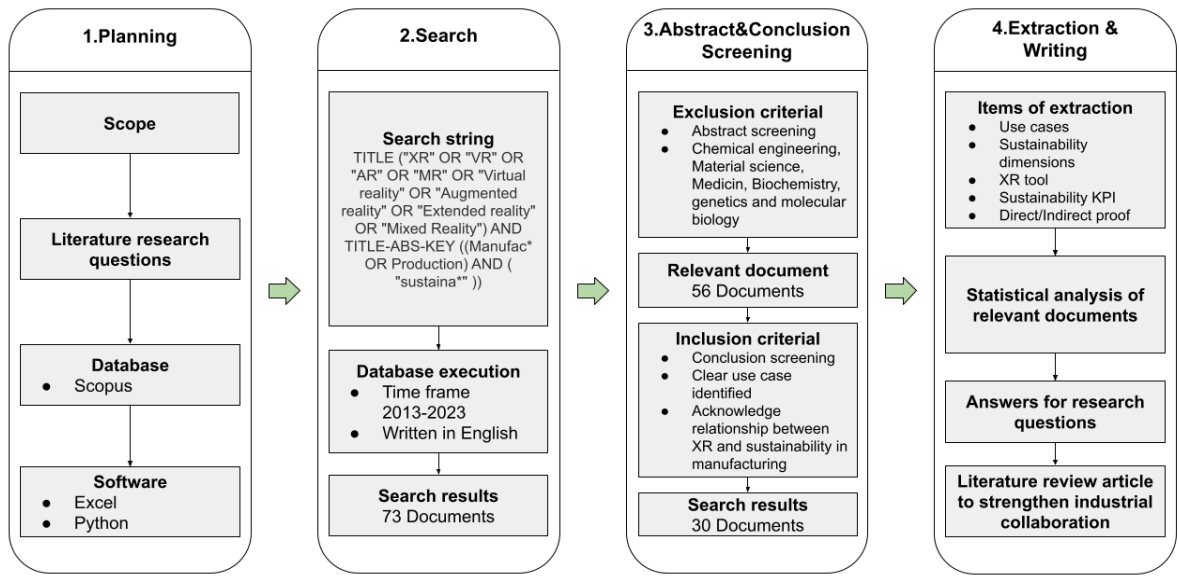

Fig. 2: Illustration of the methodology used in the literature study.

To correctly identify the corresponding sustainability pillar addressed in the papers and use-cases examined, the authors referred to the sustainability indicators defined by NIST [6][7] under each sustainability pillar as described in section 1. If the examined paper mentioned one or more indicators, tied to a specific sustainability pillar, as the result or impact of an XR implementation or use case, this paper would be categorized under corresponding pillar. This meant that the same paper could be categorized under more than one pillar and at the same time, papers not specifically mentioning a metric or outcome that could be tied to any of the indicators as suggested by NIST, would not be mapped towards any of the sustainability pillar. By mapping the recorded sustainability indicators and sustainability pillars towards the described application of XR proposed by the examined papers, clear examples and use cases could be identified that proposes environmental impact and sustainability as the result of XR technology within manufacturing.

Furthermore, to determine the focus and main aim of the papers, the distinction between the directly affected indicator of the proposed use case and indirect affected indicator as the result of a use case was made.

- **Direct impact** - This is based on quantitative evidence directly cited by the paper.
- **Indirect impact** - This is derived from qualitative claims suggesting long-term benefits of applying XR technology to enhance sustainability.

## 3 RESULTS

By analyzing the 30 papers selected for this study, a total of 39 use cases of XR technology in manufacturing were identified. In the following section a deeper analysis of the 39 use cases is described and their relationship to environmental sustainability is mapped out.

### 3.1 Sustainability perspectives analysis of literature

When categorizing the papers according to the NIST framework, a clear division in the attention between the three sustainability pillars given by the examined papers can be easily identified, see Fig. 3.

As shown on Fig. 3, a total of 36 use cases in the studied literature has been identified by its authors as having a direct or indirect impact on *economic* sustainability. For instance, via reduction of processing time or via increased production efficiency with the help of XR technology to support decision makers or operators to act more quickly [9][10]. Secondly, 15 of the examined use cases are described by their authors as having a positive impact on *social* sustainability. Often as a result of improved workers health and safety or inclusion in the workforce. The examined use cases showed great examples of XR technology as a tool to support human-centric and ergonomic workplace design or as an aid for hearing impaired workers to receive visual instructions and alerts [11][12]. *Environmental* sustainability was the least referenced sustainability pillar in the examined literature and only 13 use cases were described as having a positive impact on environmental sustainability by its authors. Some examples of which included reduction of waste as a result of improved quality assurance, resource efficiency by virtual prototypes and design reviews, or by reduced travel via virtual shop floor meetings or optimized path planning using AR [13][14][15][16][17].

Furthermore, a clear distinction can be seen between the three sustainability pillars when it comes to the direct and indirect impact on the sustainability indicators. More than half of the use cases marked under the environmental sustainability pillar (7 out of 13) mention environmental indicators as indirectly impacted by the XR technology and not as a main focus or driver of the XR use case and publication. For instance, as described in paper [17], AR technology was adopted with the aim to decrease the set up and maintenance time of machinery via remote support from a technician, however, the author also highlighted the reduced need for traveling and transportation as a positive consequence of the use case. Meanwhile, a clear majority of the use cases found under the social- and economic sustainability pillar acknowledge its indicators as the direct consequence of the XR technology and the focus and driver of the publication, see Fig. 3.

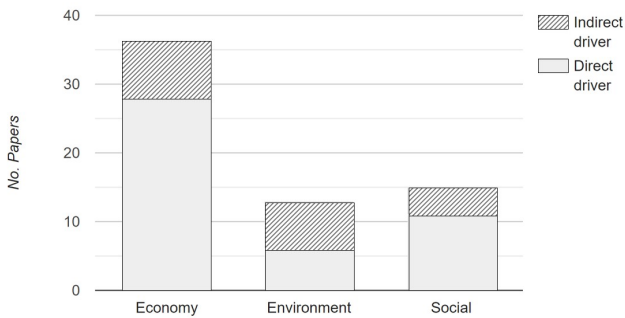

Fig. 3: Graph showing the number of papers that identifies one or more of the NIST indicators of any of the three sustainability pillars.

### 3.2  Use case categorization

The use cases described in the literature were categorized with a focus on the use and application of the XR technology as described by its authors using a deductive categorization approach. The categorization suggests cross-industry and generic use cases of XR that can be found in various areas of manufacturing such as logistics, maintenance, or operations etc. A total of seven unique categories of XR use cases were defined and are further presented as shown in Fig. 4 as well as detailed in the following paragraphs.

*Virtual prototyping*

*Virtual prototyping* uses XR as a co-creation testbed to involve customers and engineers to reduce real material for tests and prototypes. The paper [18] works on business-to-business interaction through VR for co-production, and the author addresses the potential of VR to include more parameters like energy consumption and carbon emissions with its high calculative power for sophisticated graphics. Customer involvement requires the VR demo to meet their unconscious requirement like facial expressions to capture customers' quick attitudes which are mentioned in the paper [19], it shows the virtual prototype design strategy should be respondent to customer's needs, focusing on avoiding waste and optimizing the resources.

*Production and layout planning*

The *production and layout planning* use cases manifest production cells, workstation design, and digital factory layout to avoid costly mistakes in the design e.g., collisions, etc. In paper [10], the author clarifies by using Visual Component to visualize a production cell in VR would be able to verify the layout planning of the cell. Paper [20] tests the concept of using AR based on CAD software e to visualize production cells and workstations on top of a physical layout drawing. Paper [21] gives a clear design strategy on energy and material flow dynamic visualization, and paper [22] mentioned measuring and disclosing accounting for virtual assets used in R&D activities could be another design strategy.

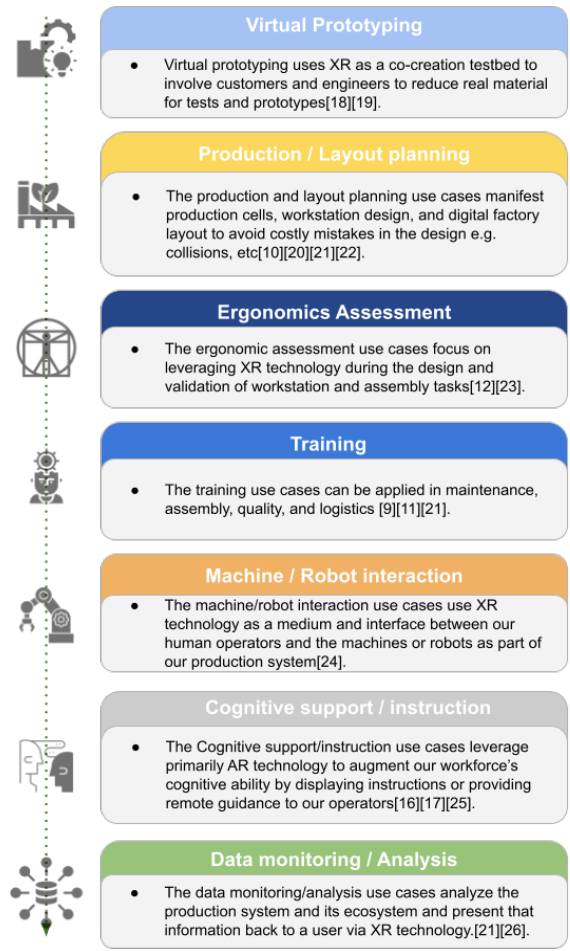

Fig. 4: Summary of the seven XR use case categories identified in literature studies.

*Ergonomic assessment*

The *ergonomic assessment* use cases focus on leveraging XR technology during the design and validation of workstation and assembly tasks. By virtually recreating a manual operation or workstation, engineers and researchers can leverage VR and AR technology to physically replicate the task in the virtual world to assess its ergonomic impact and grading. This is often with the aim of finding, and redesigning, tasks and stations that potentially pose a health and safety risk for our workforce. Often this is done in combination with a motion capture system, as seen in paper [12], or in combination with digital human modeling software and a digital manikin, such as Siemens Jack as presented in paper [23].

*Training*

The *training* use cases can be applied in maintenance, assembly, quality, and logistics [9]. Paper [9] shows the evidence that in their use case to compare training media through VR/AR and manual, which demonstrates the lowest operation speed with VR, followed by AR, which is better than manual instruction. Other publications such as [11], utilizes XR training to aid hearing impaired workers in an assembly procedure. The training design strategy should be tested and iterated for inclusiveness, especially the elderly [9] for equal virtual environment tolerance. Another design strategy mentioned in paper [21] points out the concern about raising environmentally related issues in training.

*Machine/robot interaction*

The *machine/robot interaction* use cases use XR technology as a medium and interface between our human operators and the machines or robots as part of our production system. These use cases promote human-machine interaction often by the projection of machine commands or interfaces towards the machine operator using an AR headset. As seen in paper [24] AR technology is used to visualize relevant diagnostic information and trajectories of machines and robots to its operator, thus acting as decision support with the aim to increase efficiency and workers' safety.

*Cognitive support/instruction*

The *cognitive support/instruction* use cases leverage primarily AR technology to augment our workforce's cognitive ability by displaying instructions or providing remote guidance to our operators. These use cases often promote increased quality assurance, workforce inclusiveness and ramp-up time for new operators by providing on-the-job guidance and work instructions via an AR headset, for instance as described in paper [25] and [16]. Furthermore, AR technology and cognitive support functions could be used to minimize maintenance time and travel by the possibility of remote assistance from a technician or manager via an AR application, as described in paper [17].

*Data monitoring/analysis*

The *data monitoring/analysis* use cases analyze the production system and its ecosystem and present that information back to a user via XR technology. Information on the well-being of the production system such as key measurements and KPI's can be displayed to an operator using and AR headset to support his/hers decision-making. As presented in paper [21], digital twins of facilities and production systems together with an AR headset can be used to visualize energy flows and emissions of production facilities or systems. Another use case is presented in paper [26], which showcases how a thermal camera mounted on an AR headset help operators avoid potential hot and harmful surfaces to improve workers' safety, furthermore, this technology could also be used to identify heat or energy loses in the system as well as wear and tear on machines and equipment.

### 3.3 Sustainability coverage by use case

By mapping the use cases from the literature according to the categorization described above and the previously made categorization per sustainability pillar, clear trends can be seen on how the different categories of use cases impact different pillars of sustainability. Certain categories of use cases can be directly pinpointed and linked to having a potential positive environmental impact, which provides an answer to RQ1 as well as a first direction for RQ2 as posed in this paper.

Fig. 5 showcases that five use cases from the literature of Cognitive support/instructions, as well as four use cases of Virtual prototyping have reported a potentially positive impact on environmental sustainability.

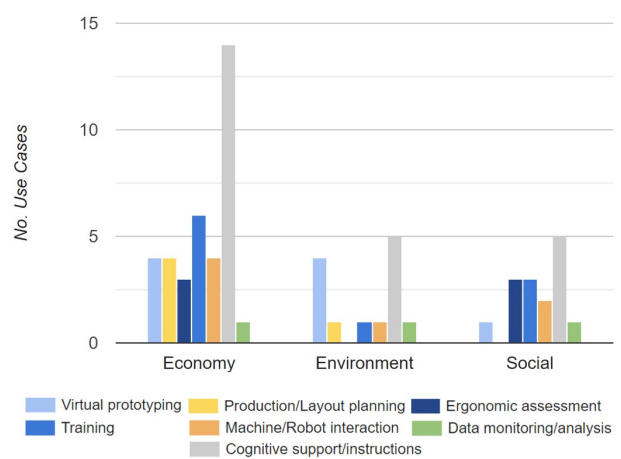

Fig. 5: Graph showing the number of particular use cases addressed by the literature related to each sustainability dimension.

### 3.4 Environmental impact by use case

To better understand how the papers and use cases identified under the environmental sustainability pillar proposes a potential positive contribution to the environment, an overview of the use cases and proposed environmental impacts was created. In Fig. 6 a summary of potential environmental impact per use case category can be seen with reference to relevant literature.

### 4 DISCUSSION

**How has sustainability been addressed in research efforts on XR for manufacturing applications?**

The results of the literature study show that there are strong relationships between sustainability and XR technology in manufacturing being addressed in the literature. The identified XR use cases from section 3 provides an overview of potential areas and applications that have documented benefits for sustainability, thus providing insights and answers to **RQ1** posed by this paper. However, this study shows that environmental sustainability is more seldomly the driver or direct indicator behind an implementation or adaptation of XR technology in manufacturing, when compared to the other sustainability pillars. Environmental sustainability is to a larger extent acknowledged as an indirect consequence of the implementation (if at all) rather than a driver. Meanwhile the results show that the majority of examined literature addresses the relationship between economic sustainability and XR in Manufacturing. Indicators and KPIs reporting on financial targets and economic values have traditionally steered the manufacturing sector [6], thus the findings in section 3 are to be expected. However, by promoting clear use cases and publications addressing the environmental impact of XR technology within manufacturing, the authors of this paper aim to challenge the view of financial KPIs as the driver of XR technology. We have shown that XR technology can have a positive impact on environmental sustainability and could be seen as a valid tool or aid to lower the environmental impact in a manufacturing system. By using referenced publications and use cases as inspiration, companies and researchers could find ways to keep developing and leveraging XR technology for their specific use case toward more sustainable manufacturing practices.

**What environmental benefits have been obtained in use cases of XR in manufacturing?**

As shown on Fig.6, use cases underscore XR's invaluable role in intertwining technological advancements with environmental sustainability goals.

*Virtual Prototyping:*

-to minimize resource consumption, waste and unnecessary travel from physical prototyping through VR cases [13][18][19].

*Production and Layout Planning:*

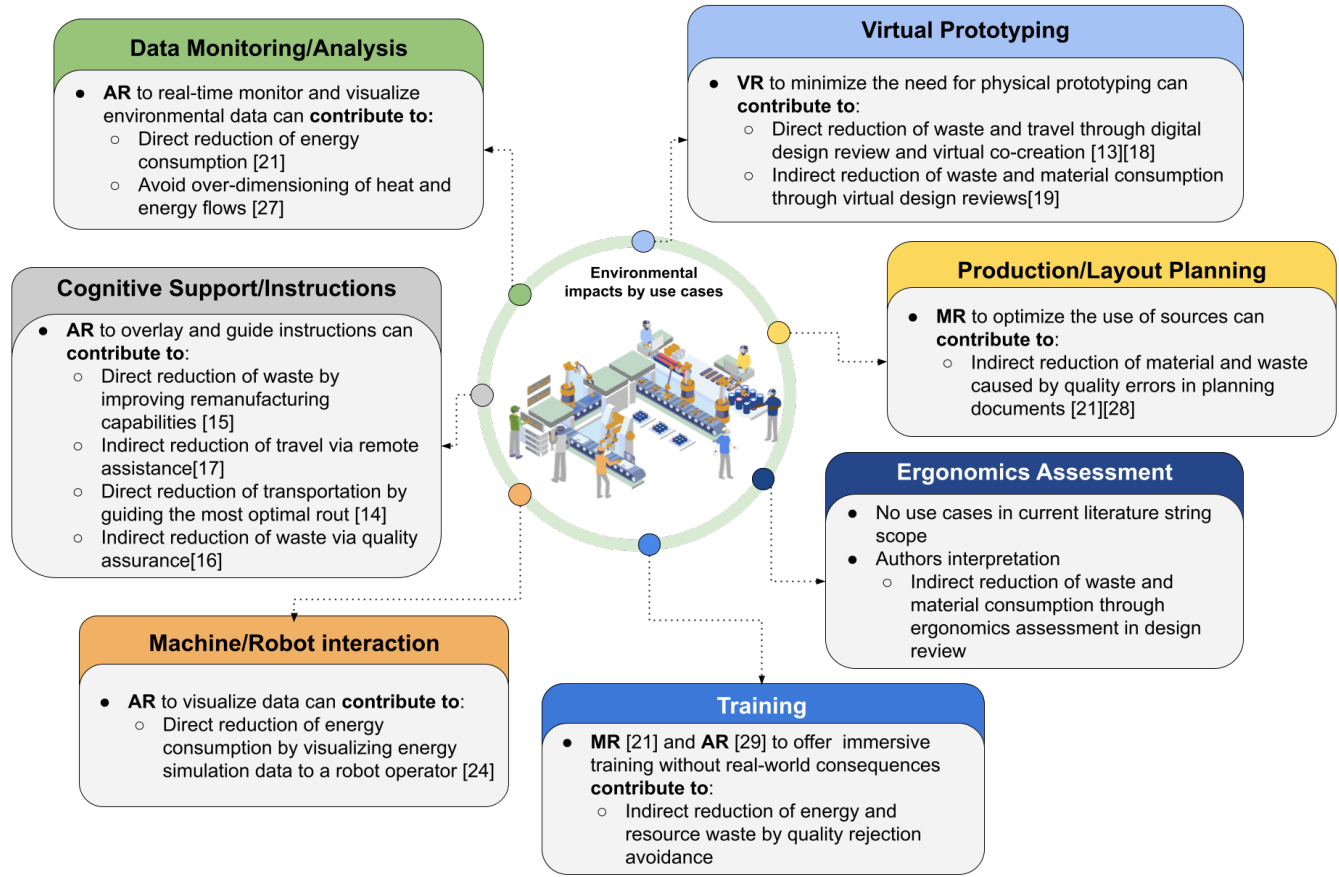

Fig. 6: Illustration showing summary of environmental impacts collected from the use cases listed in Fig.5

-to optimize, ensure efficient use of resources and reduce environmental footprints through MR cases [21][28]. This also reduces costly errors that might have environmental repercussions.

*Ergonomic Assessment:*

-to assess ergonomic factors, ensuring healthy and productive environments through MR case [30] and AR cases [23]. A sustainable workspace isn't just about eco-friendliness; it's about the well-being of workers.

*Training:*

-to offer immersive training without real-world consequences. This reduces resource use and waste, aligning with sustainable and pedagogical practices while emphasizing eco-friendly methodologies through MR [21] and AR [29].

*Machine/Robot Interaction:*

-to facilitate harmonious human-machine interactions by visualizing energy simulation data, optimizing energy and material usage through AR [24].

*Cognitive Support/Instruction:*

-to overlay and guide instructions to ensure tasks are executed efficiently and correctly the first time to minimize waste via AR [14][15][16][17].

*Data Monitoring/Analysis:*

-to real-time monitor and analyze environmental data and enable timely decisions that prioritize the planet's well-being, helping industries pivot towards greener processes via AR [21][27].

In summary, XR acts as a potential bridge to balance between human endeavors and environmental preservation. But as interpreted from Fig. 6, AR and MR seem to be more impactful than VR for environmental sustainability from the use cases identified in the research. AR and MR can guide on-site and real-time data that could more directly raise awareness of the environmental impact by overlaying critical digital data onto the actual production environment. From the use cases identified by the research string, AR and MR show more potential to take environmental sustainability as a main driver and contribute to the environmental sustainability with both quantitative and qualitative impact.

### 4.1 Limitations

This study addressed papers clearly acknowledging indicators categorized under the environmental sustainability pillar, as defined by NIST [6], use cases and papers with indirect effects on environmental sustainability that were not acknowledged by its original author have not been take into consideration. Furthermore, the delimitation of the search string and criteria used for this exploration also affects the number of publications and use cases addressing environmental sustainability found by the authors. The search string used for this study emphasized the use of XR terminology in the title of the examined literature. By expanding this search to also include papers mentioning XR terminology in the abstract, the numbers of papers would increase to 422 before exclusion and abstract screening. However, this would also introduce a majority of papers with non-concrete XR use cases and applications, hence the decision was made to keep the XR terminology in the title as a requirement for the search. By looking at other industries and sectors such as e.g., agriculture, construction, or mining, other use cases might have been found that could have been relevant for cross-sectoral learnings for manufacturing companies. Likewise, expanding the search string to include more synonyms of environmental sustainability, could have presented us with other use cases of relevance. Therefore, the referenced publications should be considered as sample of use cases that have documented a positive impact on environmental sustainability.

### 4.2 Future research directions

The future research based on the study could be the qualitative or quantitative assessment of the impact of each use case, and thus provide suggestions for the companies to prioritize the use cases for impact on sustainable manufacturing. Another parameter interesting to extract from the literature is the technical readiness level (TRL) which has a feasibility trend mapping for different use cases.

For the theoretical part, the limitation analysis of XR and different use cases are needed to see the research gap in the literature study on environmental sustainability issues, and it will help us to answer the important question: why is environmental sustainability receiving less attention in the development of XR from manufacturing? Thus, we could list the following research questions that would be potential for further research on this topic:

**RQ1**: How can XR support various indicators of environmental sustainability such as energy flow, material flow, noise and water pollution etc.?

**RQ2**: What are the current gaps in the research between XR technology and environmentally sustainable manufacturing?

**RQ3**: What are the potential negative impacts that XR brings to environmental sustainability?

## 5 CONCLUSION

This paper shows that XR technology has the potential to drive a positive impact on environmental sustainability in a manufacturing system. Thus, the objective of the paper to highlight and eliminate the gap between XR and the direct/indirect relationship with environmental sustainability has been met. To highlight the potential XR could bring to environmental sustainability and to act as inspiration for industry and researchers to continue to work together for developing XR applications for a sustainable future, this paper presents a summary and examines seven categories of potential use case to adopt, from the identified use cases of the literature study.

### 5.1 Contribution to theory

The seven XR use case categories are concluded from the review to provide a visual representation of the XR development for sustainable manufacturing. It also details pointing out the direct/indirect impact of XR use cases on environmental sustainability, and thus a potential research direction to raise awareness of environmental sustainability with XR is identified.

### 5.2 Contribution to practice

The cross analysis of the use cases to sustainability and seven categories could provide companies with toolboxes, cases and design strategies when a new case study is identified. The research also claims that the XR use cases should take sustainability principles into consideration and corresponding KPI and evaluation should be followed to direct XR for sustainable manufacturing.

## 6 ACKNOWLEDGEMENT

The authors would like to thank the Swedish innovation agency Vinnova for their funding of the PLENUM project, grant number: 2022-01704. The work was carried out within Chalmers' Area of Advance Production. The support is gratefully acknowledged.